# Optical characterization of active Photon Cages

R. Artinyan[1,*], A. Benamrouche[1], C. Belacel[1], M. Kozubova[2], A. Berthelot[2], A.M. Jurdyc[2], G.Beaudin[3], V.Aimez[3], P. Rojo-Romeo[1], J.L. Leclercq[1], R. Peretti[1], X. Letartre[1], S. Callard[1]
*1 : Université de Lyon, Institut des Nanotechnologies de Lyon (INL), UMR CNRS 5270*
*2 : Université de Lyon, Institut Lumière Matière (ILM), UMR CNRS 5306*
*3 : Institut Interdisciplinaire d'Innovation Technologique (3IT), Université de Sherbrooke, 3000 Boulevard Université, Sherbrooke, J1K OA5, Québec, Canada*

** : contact : remy.artinyan@ec-lyon.fr*

## ABSTRACT

Recently, we developed a new family of 3D photonic hollow resonators which theoretically allow tight confinement of light in a fluid (gaz or liquid): the photon cages. These new resonators could be ideal for sensing applications since they not only localize the electromagnetic energy in a small mode volume but also enforce maximal overlap between this localized field and the environment (i.e. a potential volume of nano-particles). In this work, we will present numerical and experimental studies of the interaction of a photon cage optical mode with nano-emitters. For this, PbS quantum dot emitters in a PDMS host matrix have been introduced in photon cages designed to have optimal confinement properties when containing a PDMS-based active medium. Photoluminescence measurements have been performed and the presence of quantum dot emitters in the photon cages has been demonstrated.
**Keywords:** Near-infrared microcavity, photonic-crystal, air-mode, colloidal quantum dots

## 1. INTRODUCTION

The general approach to achieve strong confinement of photons consists in high index contrast structuring of space at the wavelength scale. Numerous configurations have been used demonstrating the storage of light during a long time (high Q factor) into microcavities (small mode volume), leading to high Purcell effect. In particular, design and fabrication of multidimensional architectures for functional optical devices are of great interest for bio and health applications (such as sensing, fluorescence imaging) where it is aimed at strengthening the interactions with biological molecules or nanoparticles. For these sensing operations, hollow 3D resonators are ideal since they not only maximize the 3D light confinement but also enforce maximal overlap between the localized field and the environment (i.e. a potential volume of nano-particles). Recently, we proposed an original photonic concept to design new 3D photonic hollow resonators which theoretically allow tight confinement of light in vacuum or in a fluid (gas or liquid). The idea is to use a very thin and highly reflective membrane made of a 1D or 2D photonic crystal in semiconductor material as a non-absorbing broadband mirror. This effect, which has been demonstrated for 1D Fabry-Pérot cavities is the keystone of the resonator principle. In previous work, we focused on Si/air PC membranes. Simulations showed that reflectivity properties of these Photonic Crystal Membranes (PCM) are well preserved with curvature radius up to a few µm. We have therefore proposed to "bend" the photonic membrane in air, in order to achieve cylindrical optical cavities. Numerical simulations showed that for a 6µm diameter microcavity of 10µm-high silicon cylindrical pillars (diameter 0.27µm), strong confinement of light in air is obtained in the central hollow part of the cylinder (see Fig. 1) for relatively high Q modes (~1400 at 1560nm)[1,2].
In this work, we propose to exploit these properties to enhance light-matter interaction by coupling an active material to the optical modes of the photon cage. For this we investigated the case of a photon cage filled with a layer containing PbS quantum dots. Such emitters can be inserted into a PDMS matrix deposited inside a photon cage. The photon cage properties, though, are strongly depending on the contrast index between the pillars and the background medium. We therefore designed new photon cages adjusted to the refractive index of the PDMS medium.

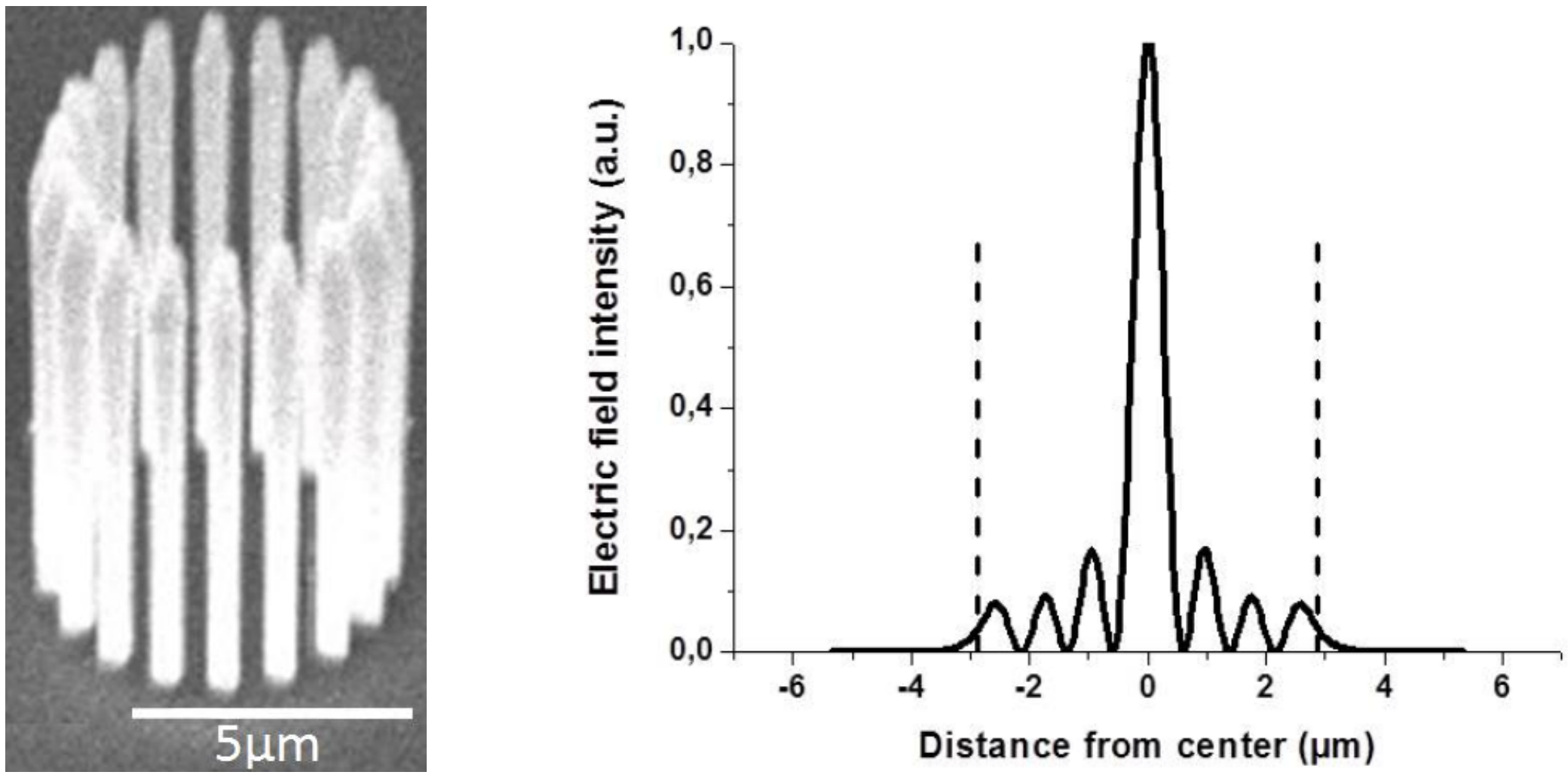

Fig. 1 : Left : SEM image of a 18-pillar, 6µm diameter photon cage. The pillars are 10µm high and their diameter is 0.27 µm; Right : Electrical field magnitude along a diameter of the cage (the dots are located on the boundaries of the photon cage) at λ=1560nm (mode quality factor Q=2500) .

In this paper, we will present the designing process of active photon cages, the activation (insertion of an active medium inside the cage) protocol and the first characterization results.

## 2. DESIGN OF ACTIVE PHOTON CAGES

To preserve the quality factor of the mode in the wavelength range around 1.5µm for a Si/low index medium cage, the geometrical parameters of the cage need to be optimized. In our case, the low index media is a PDMS resin in which variable concentration of PbS colloidal quantum dots have been inserted. The first step consists in designing the reflective membrane based on a 1D photonic crystal that allows the best reflecting properties; The period and filling factor of the PCM are then adjusted to optimize the properties of the 2D photon cage .The last step involves optimizing the photon cage diameter (or number of pillars) to obtain the best possible confinement with 10µm-high silicon pillars (light confinement in photon cages increases with the height of the pillars [2], and 10µm is the maximum height achievable with our reactive ion etching process while maintaining sufficient accuracy on the other parameters during fabrication). The photon cage designing process was conducted by 2D and 3D-FDTD simulations using the Lumerical and Harminv (by S.G.Johnson) [3] softwares.

### 2.1 Design of a Si/low index medium 1D-PC mirror

In order to obtain mirrors with high reflectivity (>0.995) over a broad wavelength range (1500-1650 nm), the PCM parameters are optimized with a refractive index contrast between PDMS (n~1.5) and silicon (n=3.48). We simulated the PCM by defining one period of the photonic crystal, using periodic Bloch conditions at the boundaries. As the basic pattern of the photonic crystal, we chose ellipses instead of circles to be able to adjust its geometry more precisely and still be able to build the designed structures.

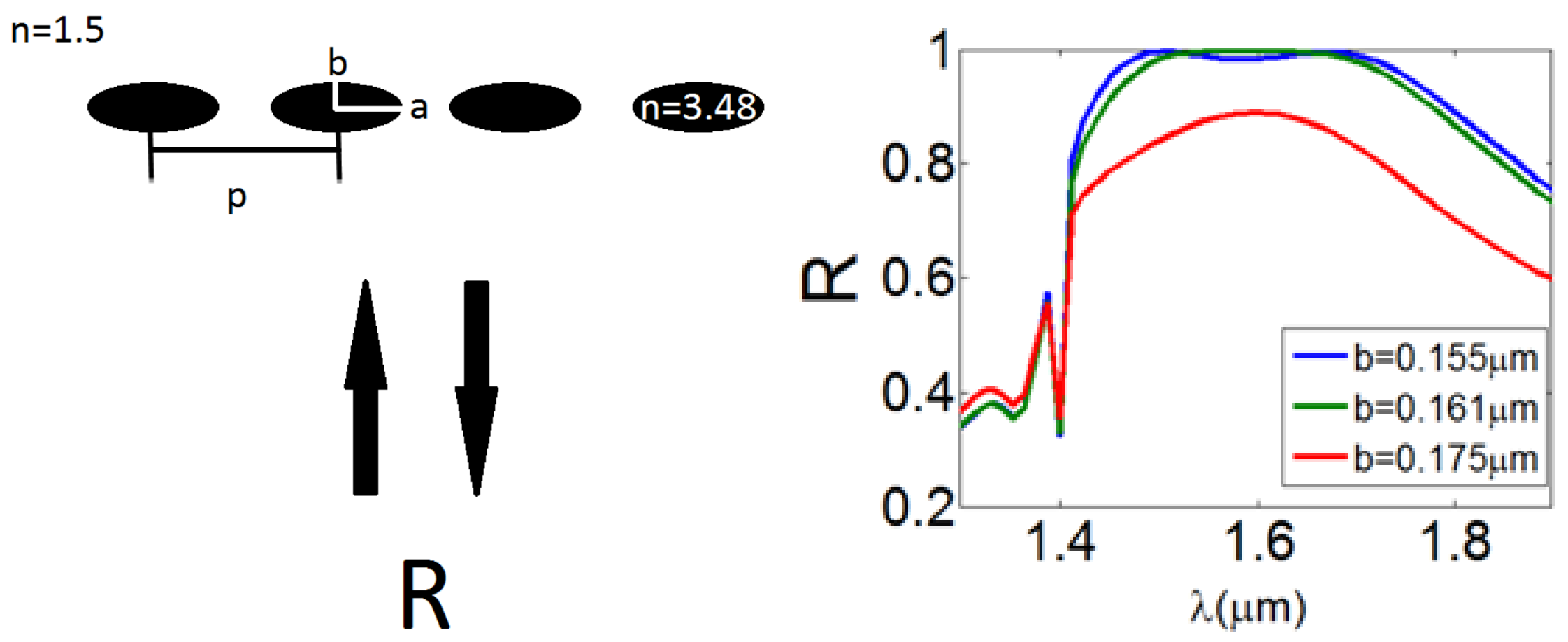

Fig. 2 : Left : simulated PCM. Right : reflectivity obtained for p=0.88µm, a=0.142 µm, and b=0.155 µm, b=0.161µm and b=0.175µm. The green curve shows good reflectivity on a broad wavelength band.

We optimized the geometrical parameters of the PCM : the ellipse major and minor radiuses a and b, and the photonic crystal period p (see Fig. 2), and obtained a broadband mirror with a reflectivity close to 1 over a wavelength range of more than 150nm. Optimal reflectivity plateau was achieved for p=0.88µm, a=0.142 µm, and b=0.161µm.

## 2.2 Two-dimensional photon cage design

The previously obtained PCM was then used as the basis for a 2D-cavity (see Fig. 3). Parameters a, b and p were first adjusted to obtain the best possible light confinement for a given 39-pillar photon cage. We obtained results quite close to the parameters of the optimal 1D-mirror : p=0.88µm, a=0.153 µm, and b=0.159µm. It can be observed that the ellipses defining the pattern of the PCM present very small eccentricity. In the case of the 2D photon cage, the numbers of pillars (or the diameter of the cage) must also be optimized to obtain the best quality factor, leading to the best light confinement.

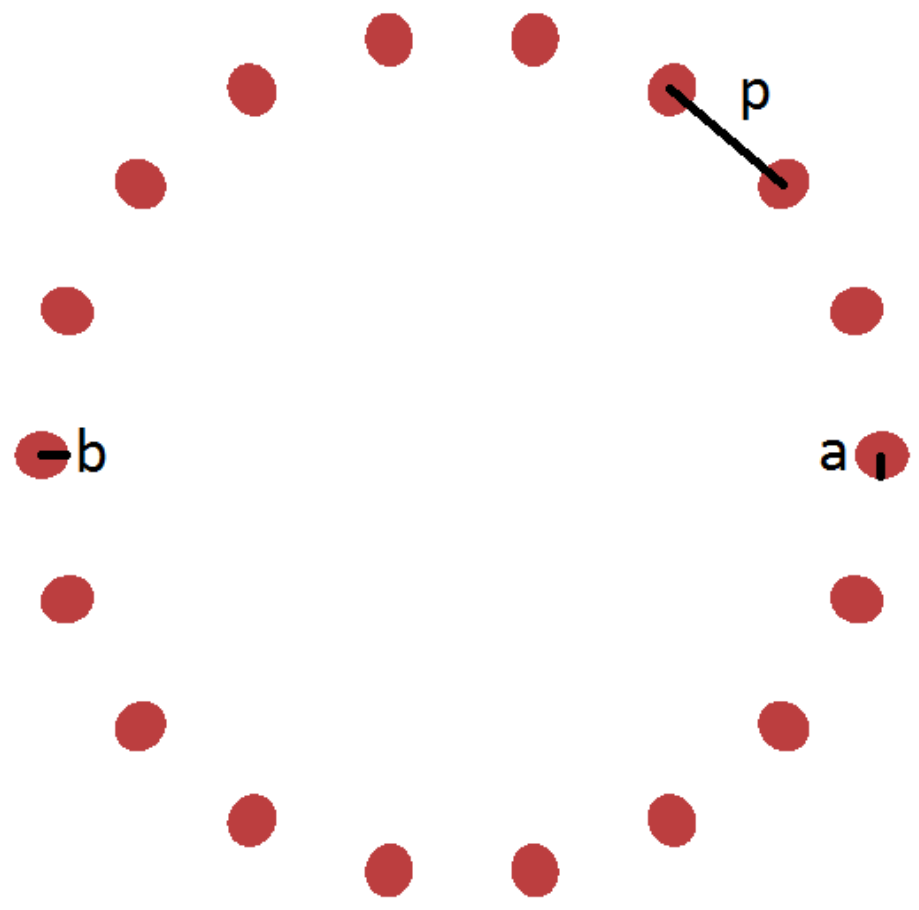

Fig. 3 : Cross-section of a 2D photon cage with 18 pillars, p=0.88µm, a=0.153 µm, and b=0.159µm. The pillars are infinitely high.

We then proceeded to investigate the influence of the number of pillars (or photon cage diameter). We simulated the response of the photon cage to an excitation by a broadband dipole (center wavelength 1550nm, width 200nm) polarized along the axis of the cylindrical cavity and located at its center. The highest quality factor of all modes in a photon cage is overall increasing with the number of pillars of the cage, and it reaches ~21000 for a 76-pillar (or 21.3µm diameter) photon cage. The wavelength variation of the mode with the highest quality factor in a cage as a function of the number of pillars of the cage was also studied. It increases with the number of pillars up to 1613nm, and then drops to 1517nm and starts increasing again. These results are presented in Fig. 4.

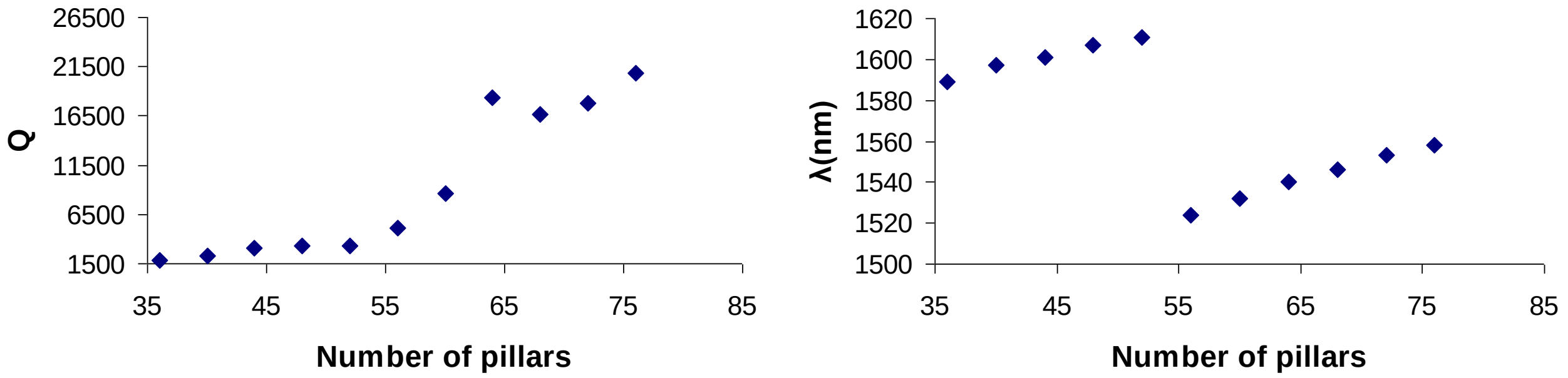

Fig. 4 : Left : best mode quality factors of the optimized 2D-photon cage against the number of pillars. Right : wavelength of mode with the highest quality factor for cages of given numbers of pillars.

The constant increase of the mode wavelength and quality factor with the photon cage diameter is similar to the behavior of a Fabry-Pérot resonator. The sudden drop in the mode wavelength is explained by the optimal reflectivity plateau of the PCM. Indeed, as long as its wavelength stays on the PCM optimal reflectivity plateau, the mode with highest quality factor retains the same number of radial orders. Then, as the PCM is no longer efficient at the mode wavelength, the quality factor of another mode with more radial orders becomes higher than that of the previously best mode. The wavelength of this new mode increases again starting from the bottom end of the PCM optimal reflexion range (see Fig. 5).

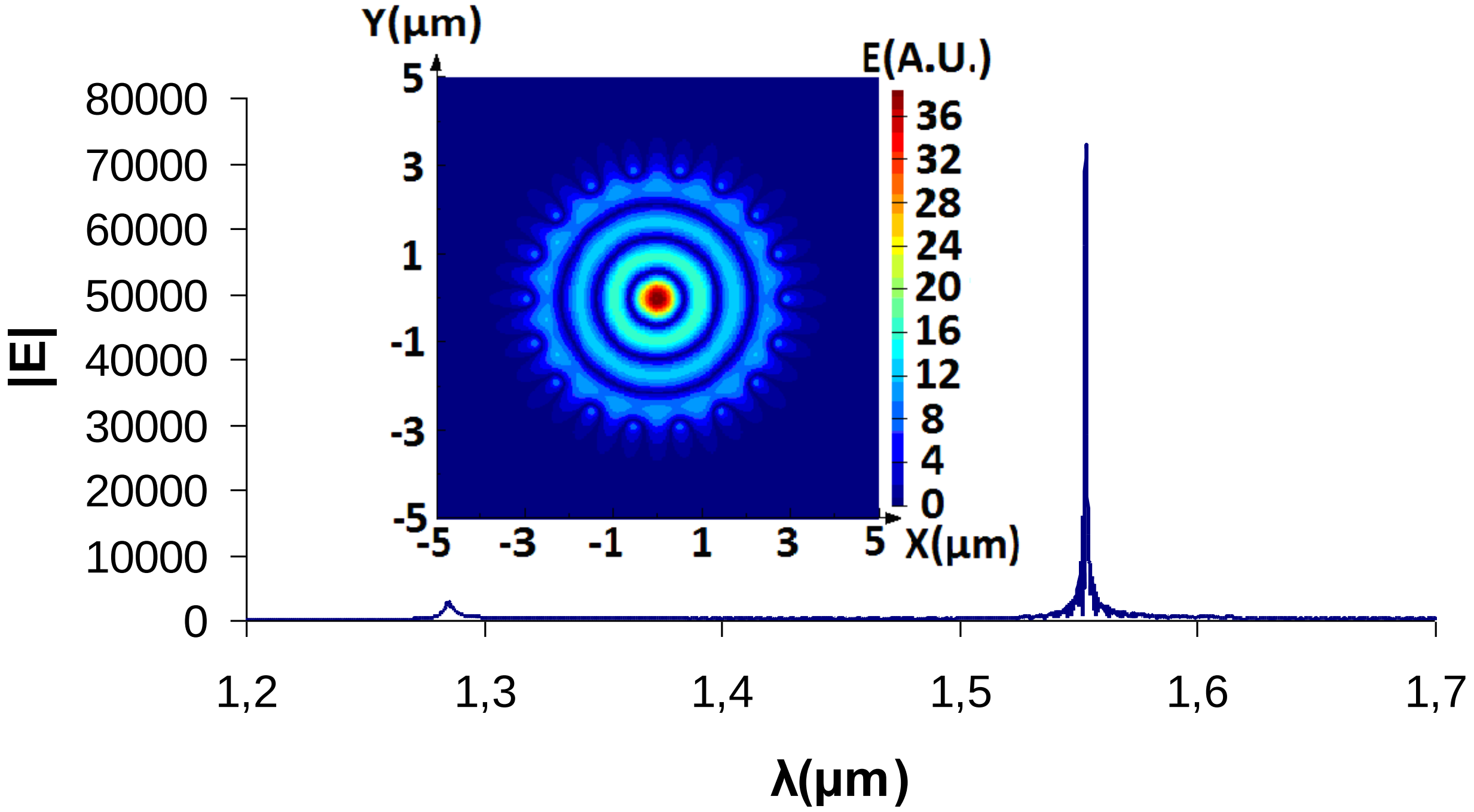

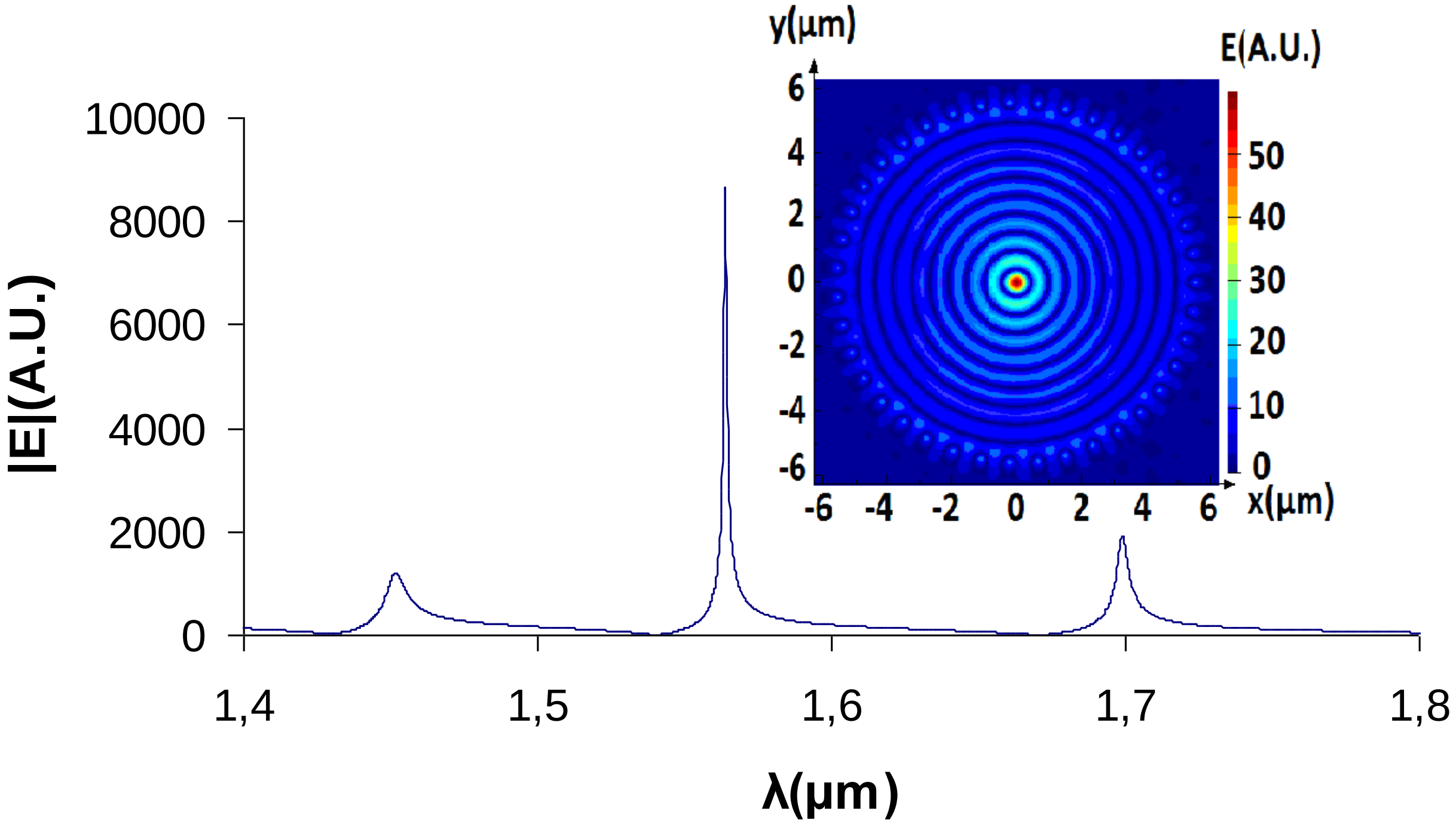

Fig. 5 : Top : Electrical field magnitude in the 18-pillar photon cage described in Fig. 1. Insert : |E|-field mapping at λ=1560nm (Q~50000). Bottom : Electrical field magnitude in the 39-pillar photon cage with p=0.88µm, a=0.153 µm, and b=0.159µm. Insert: |E|-field mapping at λ=1564nm (Q~3100).

## 2.3 Three-dimensional photon cage

The work realized for 2D photon cages only accounts for radial losses of the structures. We also simulated 10µm-high cages from the 2D-parameters obtained in the previous part (see Fig. 6). The 3D quality factor reaches a maximum for a 64-pillar cage (diameter 17.9µm ; Q>2600) To estimate the losses in the third dimension, we also plotted the difference between the losses in the 3D-cage (~$1/Q_{3D}$) and the losses of the 2D-cage(~$1/Q_{2D}$) . The difference should represent the losses in the vertical direction $\Delta = (1/Q_{3D})-(1/Q_{2D})$. For a mode with a given number of radial orders, the losses in the third dimension increase with the cavity diameter. Indeed, 2D losses become negligible compared to losses in the third dimension. A 64-pillar photon cage has therefore been chosen as a suitable candidate for fabrication and activation.

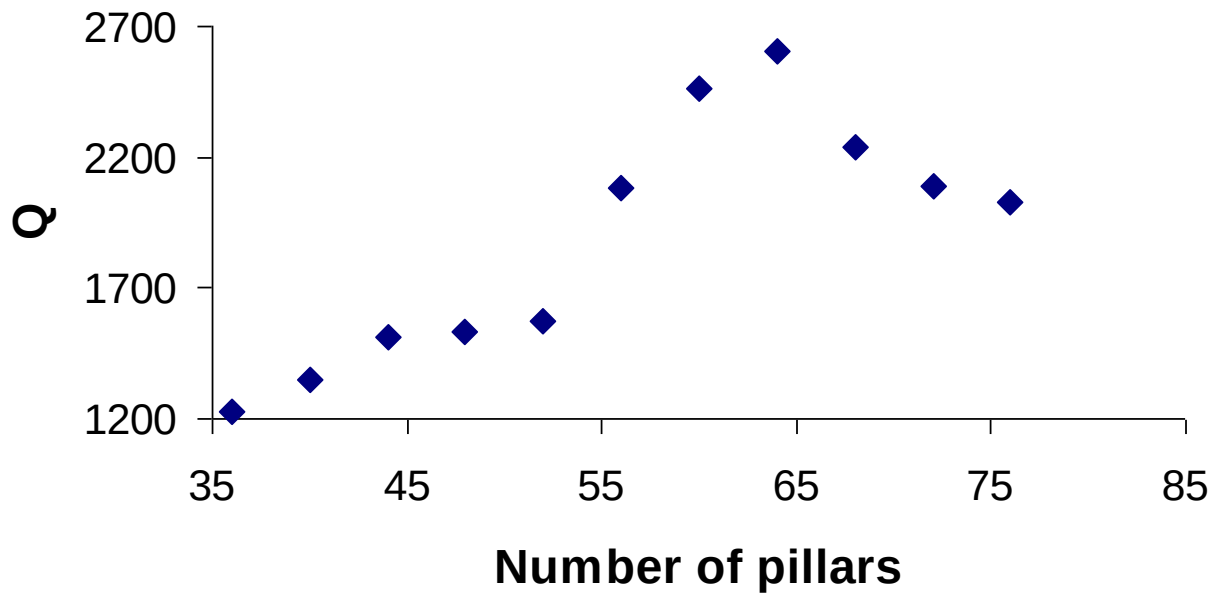

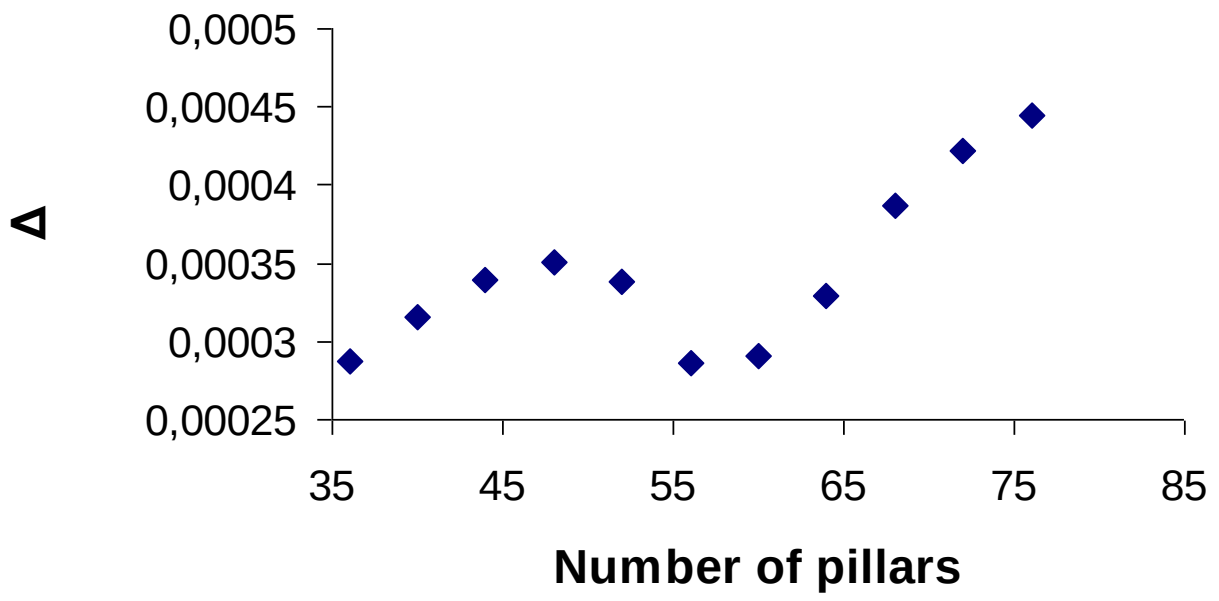

Fig. 6 : Left : best mode quality factors of the optimized 3D-photon cage against the number of pillars. Right : 3D losses minus 2D losses for cages of given numbers of pillars.

## 3. ACTIVATION OF A PHOTON CAGE

The challenge addressed in this part is to deposit a thin layer (10µm-high) of PDMS containing PbS quantum dots inside a photon cage. The PbS concentration in this layer needs to be homogeneous and controlled.

In order not to damage the photon cages and to obtain an homogeneous repartition of emitters inside and around the photon cages, we chose to deposit a drop of liquid solution containing the light emitters on photon cages. The PDMS is diluted in heptane, thus creating a fluid whose viscosity is small enough to be uniformly deposited on the sample containing photon cages. PbS quantum dots are then added to the solution so that a desired concentration of light emitters is reached. The result is then deposited onto the photon cages, and after drying and heating steps, the heptane is evaporated and we obtain a 10µm-high solid layer of homogeneous active medium on a big area of the sample (see Fig. 7). To prevent blue-shifting of the PbS dots caused by oxidation, we coat the active photon cage with a 10nm-thick layer of alumina ($Al_2O_3$) deposited by pulsed laser deposition (PLD).

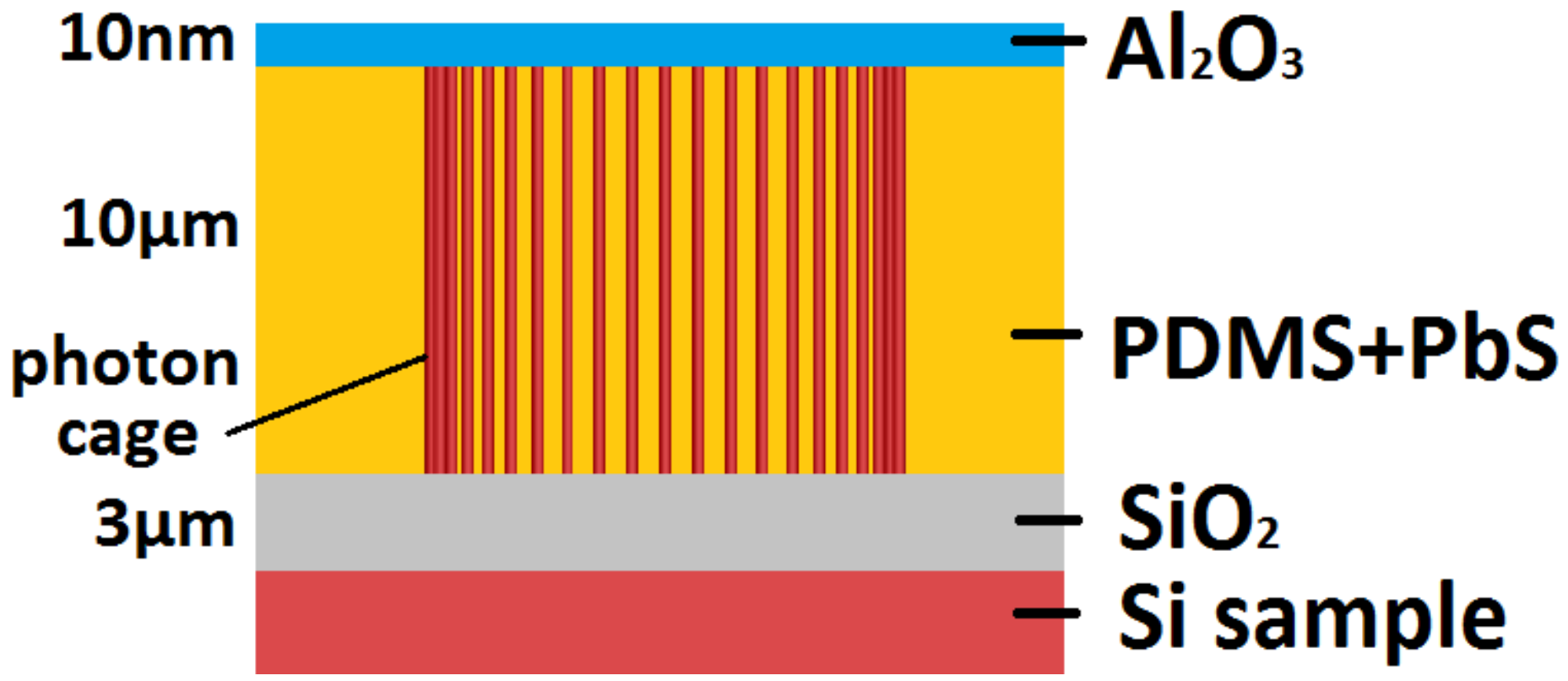

Fig. 7 : Active photon cage. The final sample covers an area of several square millimeters containing multiple photon cages embedded in a PDMS layer containing PbS quantum dots.

## 4. CHARACTERIZATION RESULTS

Photoluminescence experiments have been performed on samples like the one shown in Fig.7. The photon cages used for the first characterization attempts, though, were composed of 39 cylindrical pillars. We used the experimental setup described in Fig. 8.

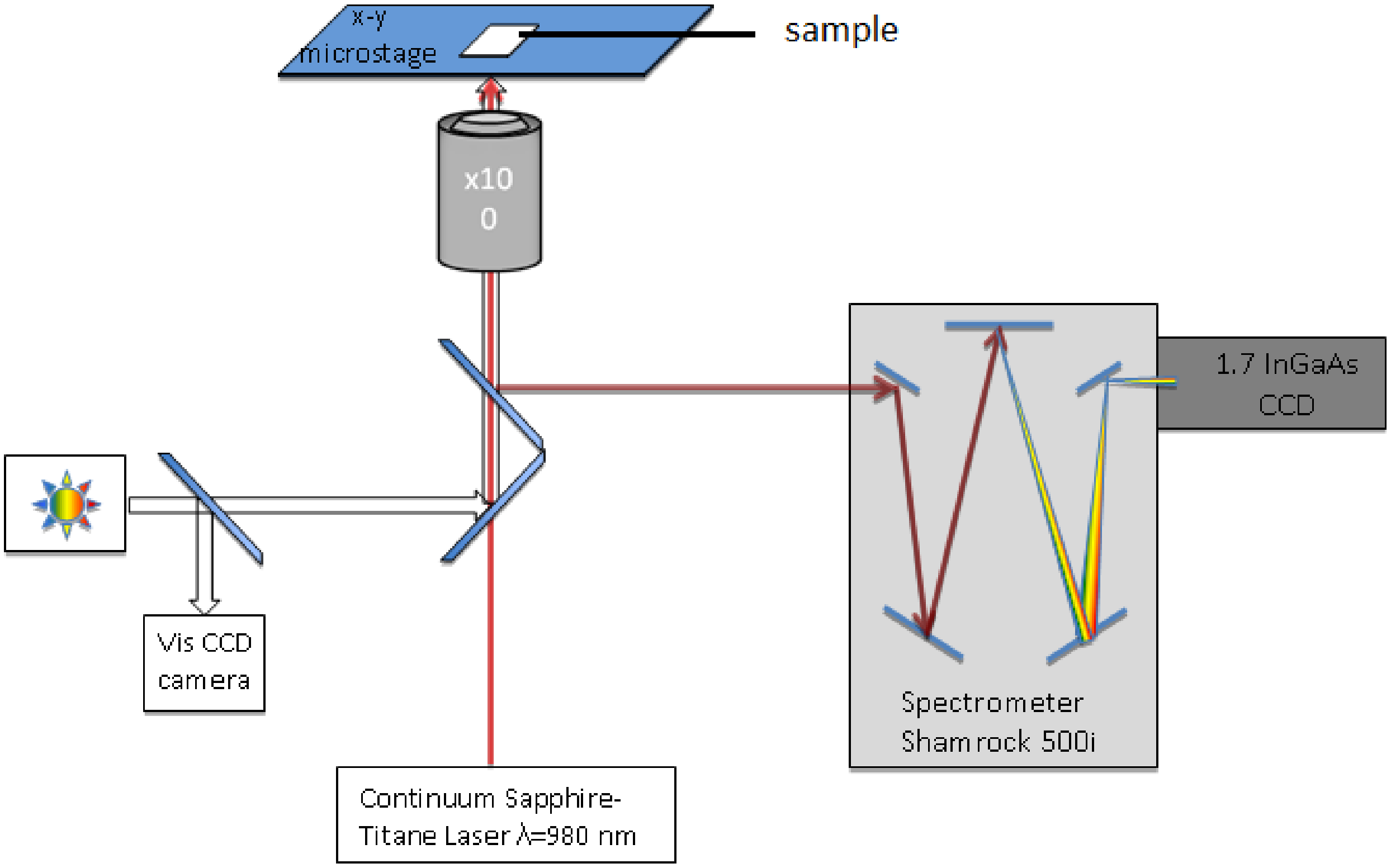

Fig. 8 : Experimental photoluminescence characterization setup. The sample is similar to Fig. 7. As a light source, we used a Ti-Sa laser at λ=980nm. The laser spot on the sample has a diameter of 2µm.

We focus a laser beam at λ=980nm on the sample and measure the photoluminescence of the PbS quantum dots for laser spots inside and outside of photon cages. The results show an enhancement of measured photoluminescence inside photon cages (see Fig. 9). The measured emission spectrum, though, is not structured by coupling with photon cage modes.

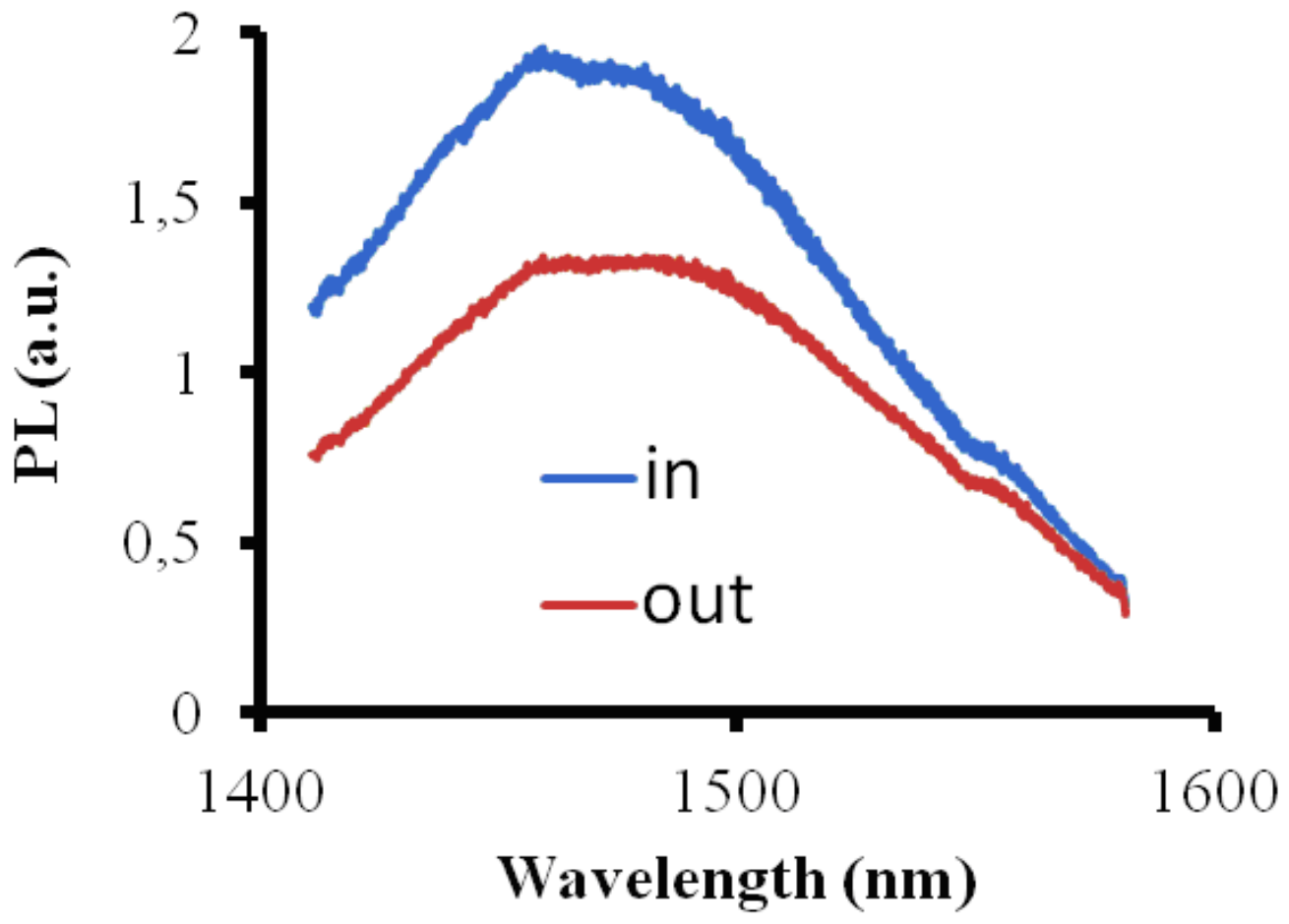

Fig. 9 : Photoluminescence of PbS quantum dots inside photon cages (blue) and outside of photon cages (red). Despite photoluminescence enhancement, the measured spectrum of the light emitters seems to be unaffected by the cavities.

One possible explanation is that the observed photon cages are non-optimized 39-pillar photon cages, and the quality factors of our photon cage modes in the range delimited by the PbS emission spectrum may therefore be quite low. This underlines the importance of the photon cage designing process described above. Still, the PbS emission spectrum measured inside of photon cages proves that we successfully inserted an active medium in our microcavities.

# 5. CONCLUSION AND FUTURE PROSPECTS

We have designed new photonic micro-cavities adapted to the refractive index contrast between silicon and an homogeneous medium containing PbS quantum dots. A process has been established to reliably insert such a medium inside photon cages without damaging them. The first photoluminescence measurements performed on active photon cages show emission from the PbS quantum dots inside of photon cages. This suggests light-matter interaction between the photon cages and an active medium. This would be instrumental in using the photon cages as sensors, for example with the insertion of a micro-fluidic channel inside a photon cage. The photon cage modes, though, are yet to be pinpointed, probably due to their weak quality factors in the observed samples. Alternative techniques to characterize them include studying the losses from photon cage modes, producing active NSOM probes or using metal-coated multimode NSOM [4] probes on active photon cages. These ways are currently being explored in our team.

# ACKNOWLEDGEMENTS

We thank L.Lalouat and B.Gonzalez-Acevedo for their help and advice regarding FDTD computing and A.Pereira for performing the pulsed laser deposition of alumina on the active cage samples. IMUST is acknowledged for the funding of this research.